# Superconductivity at 41.0 K in the F-doped LaFeAsO$_{1-x}$F$_x$

**Wei Lu, Xiao-Li Shen, Jie Yang, Zheng-Cai Li, Wei Yi, Zhi-An Ren\*, Xiao-Li Dong, Guang-Can Che, Li-Ling Sun, Fang Zhou, Zhong-Xian Zhao\***

National Laboratory for Superconductivity, Institute of Physics and Beijing National Laboratory for Condensed Matter Physics, Chinese Academy of Sciences, P. O. Box 603, Beijing 100190, P. R. China

Abstract:

Here we report the superconductivity in the LaFeAsO$_{1-x}$F$_x$ system prepared by high pressure synthesis. The highest onset superconducting transition temperature ($T_c$) in this La-based system is 41.0 K with the nominal composition of LaFeAsO$_{1-x}$F$_x$ (x = 0.6), which is higher than that reported previously by ambient pressure synthesis. The increase of $T_c$ can be attributed to the further shrinkage of crystal lattice that causes the stronger chemical pressure on the Fe-As plane, which is induced by the increased F-doping level under high pressure synthesis.

Since the discovery of high temperature superconductors about two decades ago [1], much efforts have been devoted to the searching for new high-Tc superconductors. Recently research interests are triggered by the report [2] that a layered F-doped arsenic-oxide LaFeAsO$_{1-x}$F$_x$ shows a superconducting transition temperature ($T_c$) of 26K. Following the above result, a series of new superconductors of ReTAsO$_{1-x}$F$_x$ and ReTAsO$_{1-x}$ (Re=La, Ce, Pr, Nd, Sm, Gd, and T= Fe, Ni ) were synthesized [3-9] with a highest $T_c$ of 55 K in the Sm-based system [7]. The crystal structure of these quaternary equiatomic ReTAsO is simple compared with cuprates, and can be categorized as ZrCuSiAs-type showing tetragonal layered structure with the space group P4/nmm, where Fe-As layer and Re-O layer stack alternately along the c-axis. The recent theoretical calculation shows the electron-phonon coupling is not strong enough to generate such high superconducting transition temperature in the LaFeAsO$_{1-x}$F$_x$[10]. It's also surprising that the strong magnetic element Fe may play an important role in superconductivity, which may supply some helpful information to study the puzzling origin of unconventional high temperature superconductors. Here we report the superconducting transition at 41 K in the LaFeAsO$_{1-x}$F$_x$ system.

A series of polycrystalline LaFeAsO$_{1-x}$F$_x$ samples were synthesized by high pressure (HP) method. At first, LaAs powder was prepared by La chips and As pieces, which were sintered in the muffle furnace at 650°C for 12 hours and then at 950°C for 12 hours. After that LaAs powder and Fe, Fe$_2$O$_3$, FeF$_2$ powders (the purities of all starting chemicals are better than 99.99%) were mixed together according to the nominal chemical formula of LaFeAsO$_{1-x}$F$_x$, then ground thoroughly and pressed into small pellets. These pellets were sealed in boron nitride crucibles and sintered in a high pressure synthesis apparatus at 1250 °C under the pressure of 6 GPa for 2 hours. A series of polycrystalline LaFeAsO$_{1-x}$ samples were also prepared for comparison. Compared with the ambient pressure (AP) method using an evacuated quartz tube, the HP method is more convenient and efficient for the synthesis of gas-releasing compound with super-high pressure-seal.

The resulting samples were characterized at room temperature by powder X-ray diffraction (XRD) analysis on an MXP18A-HF type diffractometer with Cu-Ka radiation from 20° to 80° with a step of 0.01°. Fig. 1 shows the XRD patterns of HP nominal sample LaFeAsO$_{0.4}$F$_{0.6}$ and undoped AP sample LaFeAsO, respectively, and the vertical bars at the bottom indicate the theoretical simulation of diffraction peaks. It's obvious that each of the comparatively strong diffraction peaks in the theoretical simulation has its counterpart in the patterns of HP nominal sample LaFeAsO$_{0.4}$F$_{0.6}$ and undoped AP sample LaFeAsO in spite of few extra weak peaks coming from impurities, which indicates the principal phase can be well indexed on the basis of tetragonal ZrCuSiAs type structure with the space group P4/nmm. Fig. 1 also shows that diffraction peaks of HP nominal sample LaFeAsO$_{0.4}$F$_{0.6}$ shift to high diffraction angle in comparison with the undoped AP sample LaFeAsO, indicating the lattice is shrunk due to sufficient F-doping by the high pressure synthesis method. The refined lattice parameter for this HP sample is a = 3.991(5) Å and c = 8.700(8) Å, compared with the AP sample (a = 4.033(5) Å and c = 8.739(1) Å), the a-axis shrinks by 1% while c-axis shrinks by 0.45%. The observed shrinkage on the Fe-As plane of this sample is the largest among all the reported LaFeAsO$_{1-x}$F$_x$ and LaFeAsO$_{1-x}$ samples, which indicates a strong inner chemical pressure on the Fe-As plane.

The electrical resistivity measurement for LaFeAsO$_{1-x}$F$_x$ samples was performed by a standard four-probe method down to 4.5K at ambient pressure with no applied field. Fig. 2 shows the temperature dependence of resistivity for the HP sample with the nominal composition of LaFeAsO$_{0.4}$F$_{0.6}$. The electrical resistivity gradually decreases when temperature decreases from ambient temperature, then drops dramatically below 41.0 K，which we refers to as the onset $T_c$, and finally becomes unmeasurable below 30 K, the transition width is about 11 K, as shown in the inset of Fig. 2. The onset $T_c$ of 41.0 K is the highest one among all the reported data of La-based compounds in the Fe-As family.

The magnetization measurement was performed on a Quantum Design MPMS XL-1 system during warming cycle under an applied field of 1 Oe after zero field cooling (ZFC) and field cooling (FC) process. The DC-susceptibility for the HP sample with the nominal composition of LaFeAsO$_{0.4}$F$_{0.6}$ is shown in Fig. 3. The sharp magnetic transition of the susceptibility curve and the large shielding signal indicate the good quality of this superconducting component. The differential curve indicates the occurrence of Meissner effect at 31 K , which corresponds to the $T_c$(zero) on the resistivity curve. The onset diamagnetic transition can be clearly observed from the enlarged ZFC curve at 40 K, as shown in the inset of left plot in Fig. 3. The enhancement of $T_c$ in this composition results from the strong chemical pressure on the Fe-As plane. Comparing with the oxygen-deficient superconductors LaFeAsO$_{1-x}$ that we previously reported [8], the F-doping samples have better superconducting properties. This is because that F-doping can help form more stable structural phase than oxygen vacancies in this La-based system due to the large atomic radius of La. The phase diagram for La-based system is different from that of Nd-based system [8] and will be further studied and reported later.

Acknowledgements:

We thank Mrs. Shun-Lian Jia for her kind helps in resistivity measurements. This work is supported by Natural Science Foundation of China (NSFC, No. 50571111 & 10734120) and 973 program of China (No. 2006CB601001 & 2007CB925002). We also acknowledge the support from EC under the project COMEPHS TTC.

Corresponding authors:

    Zhi-An Ren: renzhian@aphy.iphy.ac.cn

    Zhong-Xian Zhao: zhxzhao@aphy.iphy.ac.cn

Figure captions:

Fig. 1: The X-ray diffraction patterns of the HP nominal sample LaFeAsO$_{0.4}$F$_{0.6}$ and undoped AP sample LaFeAsO. The vertical bars at the bottom are the theoretical simulation for the diffraction peaks of the undoped LaFeAsO sample.

Fig.2: The temperature dependence of electrical resistivity for the HP nominal sample LaFeAsO$_{0.4}$F$_{0.6}$. The resistivity begins to drop dramatically below 41.0 K and the zero resistivity appears at 30 K, as shown in the inset.

Fig. 3: The temperature dependence of DC susceptibility and differential ZFC curve for the HP nominal sample LaFeAsO$_{0.4}$F$_{0.6}$. The diamagnetic signal appears at 40 K according to DC susceptibility, as shown in the inset.

Fig. 1

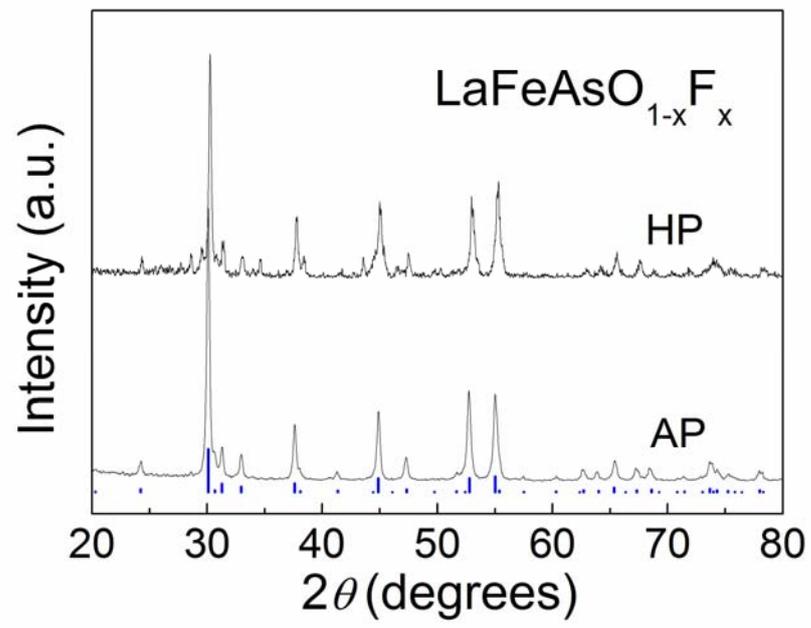

Fig. 2

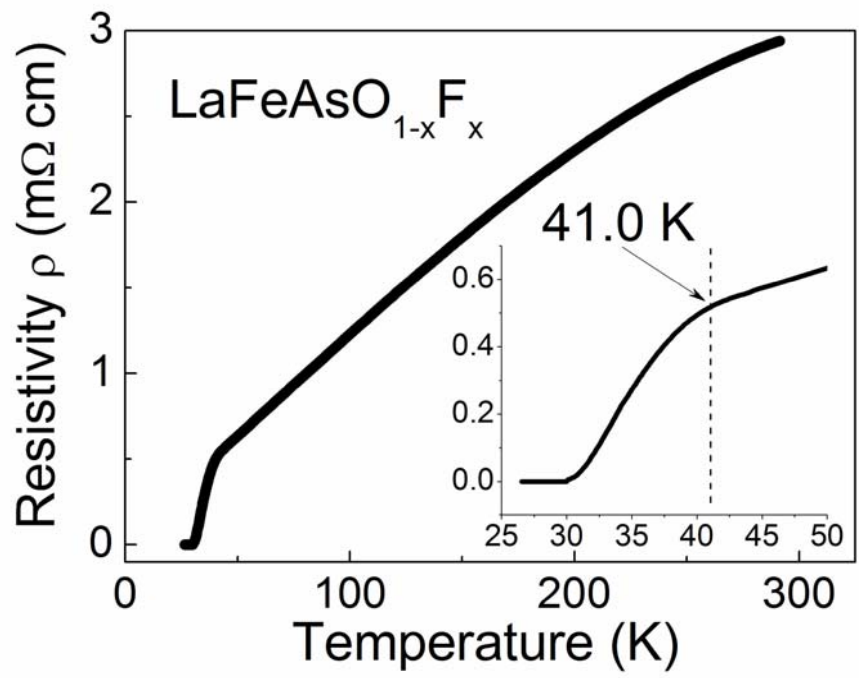

Fig. 3

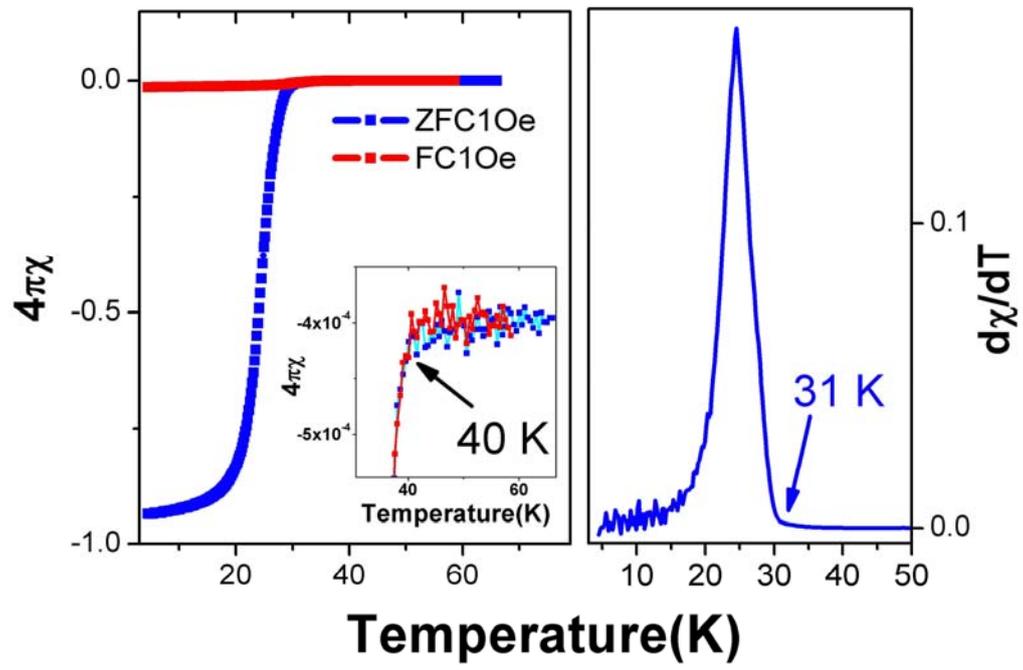